\documentstyle[colordvi,prl,aps,psfig,multicol]{revtex}
\begin{document}

\def\be{\begin{equation}}
\def\ee{\end{equation}}
\def\ba{\begin{eqnarray}}
\def\ea{\end{eqnarray}}
\title{Number-conserving rate equation for sympathetic cooling of a boson gas}
%: Information about the Growth of the Population in the Condensate Band}
%\title{Growth of a Bose-Einstein Condensate in a number-conserving system
%under Sympathetic Cooling: a Comparison with the Quantum Kinetic Theory}

%\title{Master equation for Strongly Condensed Trapped Systems under
%Sympathetic Cooling}
\author{A. N. Salgueiro$^1$\footnote{e-mail: andreans@mpipks-dresden.mpg.de}}
\address{$^1$ Max-Planck Institut f\"ur Physik komplexer Systeme, N\"othnitzer
Strasse 38, D-01187, Dresden, Germany}
%\address{$^2$Max-Planck-Institut f\"ur Kernphysik, D-69029 Heidelberg, Germany}

\date{\today}

\maketitle

\begin{abstract}
We derive a particle number-conserving rate equation for the ground state 
and for the elementary excitations of a bosonic system which is in contact
with a gas of a different species (sympathetic cooling). We use the 
Giradeau-Arnowitt method and the 
model derived by Lewenstein {\it et. al.} with an additional assumption: 
the high-excited levels thermalize much faster with the cooling agent than 
the other levels. 
Evaporation of particles, known to be important in the initial stages of the
cooling process, is explicitly included.
\end{abstract}


\begin{multicols}{2}
\section{Introduction}

The development of cooling techniques has opened up the
possibility of studying ultracold gases. In particular, the quantum
degeneracy of bosons and fermions has been investigated. Basically
there are two different processes which can be used to cool an
atomic or molecular ensemble, evaporative cooling and sympathetic
cooling. However, there are gases in which the interaction is too
weak for evaporative cooling to work. In addition, this method
fails for identical fermions at low temperatures because of the
exclusion principle ~\cite{fer}. In this case, sympathetic cooling
can be used as an alternative to evaporative cooling: a cold gas
is in thermal contact with another gas to be cooled. The first
application of the sympathetic cooling method was in the context of
the cooling of charged particles \cite{charged} and later it was
extended to neutral atoms \cite{h}. Only recently has it been
applied in the context of ultracold atoms. In particular, quantum
degeneracy of bosons and fermions has been achieved by
thermalization of atoms of the same species in different internal
states \cite{sdi}, between two isotopes of the same species
\cite{iso}, and finally between atoms of different species
\cite{dsp,allard,ketterle,matthias}. Moreover, in contrast to
evaporative cooling, sympathetic cooling does not lead to a
significant loss of either the cooled gas or the cooling agent.

%\begin{equation}
%\frac{1}{x}\mathop{\longrightarrow}_{t\rightarrow \infty}
%\end{equation}
From the theoretical point of view, classical and quantum models
have been derived to describe the dynamics of thermalization of one-species
and two-species systems. In the classical Boltzmann regime, an analytical 
formula for the
evolution of the temperature in a mixture of non-equal mass atoms and equal
mass atoms has been derived in Refs.\cite{allard,cla}. A quantum mechanical
description is given in terms of Boltzmann equations \cite{bol}
and master equations
\cite{master,pap01,qkt3,l1}.
In Ref.\cite{bol}, sympathetic cooling of an atomic Fermi gas by a Bose gas
is studied. In particular, the equilibrium temperature and the relaxation 
dynamics are obtained. In
Ref.\cite{master}, for the first time, a master equation describing 
the sympathetic cooling of a
system $A$ is derived, treating the cooling agent $B$ as a heat bath at a 
fixed temperature. More recently, by means of decoherence and 
ergodicity arguments a rate equation for sympathetic cooling 
of harmonically trapped bosons or fermions is obtained \cite{pap01}. In
Refs.\cite{qkt3,l1} a detailed quantum kinetic master equation, which
couples the dynamics of a trapped condensate to the vapor of
noncondensate particles, is obtained along with information about the growth
of a Bose-Einstein condensate. In
Refs.\cite{l2,qkt1,qkt2,qkt4,qkt5,qkt6,qkt7,l3,j1}, one can find more 
information about the growth of the condensate.

The purpose of this work is to describe the sympathetic cooling
process of a total-number-conserving system in terms of particles
in the ground state and elementary excitations. We use
the particle-number-conserving Girardeau-Arnowitt formalism
\cite{gira} and the sympathetic cooling model developed by M.
Lewenstein et al. \cite{master} with the additional assumption, that,
the highly excited levels of the trap thermalize much faster
with the cooling agent than the other levels (or in other words,
the highly excited levels of the trap are in thermal equilibrium
with the cooling agent). 
%The particle-number-conserving 
%Girardeau-Arnowitt formalism \cite{gira} is based on the
%annihilation $\hat{\beta}_0=
%{(\hat{N}_0+1)}^{-1/2}\hat{a}_{0}$  and  creation $\hat{\beta}_0^{\dagger}=
%\hat{a}_{0}^{\dagger}{(\hat{N}_0+1)}^{-1/2}$ operators 
%of one particle in the ground-state and phonon operators 
%$\hat{\beta}_{\vec n}=\hat{\beta}_0^{\dagger}\hat{a}_{\vec n}$ and
%$\hat{\beta}_{\vec n}^{\dagger}=\hat{a}_{\vec n}^{\dagger}\hat{\beta}_0$,
%where $\hat{N}_{0}$ is the number operator of particles in the ground-state,
%and $a_{\vec{n}}$ and $a^{\dagger}_{\vec{n}}$ are the usual annihilation and
%creation operators for the trap level $\vec n$. These operators obey 
%the Bose commutation relations, $[\hat{\beta}_{\vec n},\hat{\beta}_{\vec
%n'}^{\dagger}]=\delta_{\vec{n},\vec{n'}}$. 
The advantage of the 
Girardeau-Arnowitt formalism is that it covers all cases ranging from a
total absence of population in the ground state, $\textsf{n}_{0}=0$,
to a highly populated ground state $\textsf{n}_{0}\simeq N$.
The assumption of the fast thermalization of the highly excited levels 
of the trap is related to
the separation of
the levels of the trap into two different bands, in the same
way proposed by C.W. Gardiner et al. in Refs. \cite{qkt3,l1}, 
the {\it condensate} ($B_{C}$) band and the {\it noncondensate} band 
($B_{NC}$), where the latter is in
thermal equilibrium with the cooling agent. The condensate band
includes all trap levels which are directly influenced by the presence of the
condensate. The noncondensate band contains energy levels that are
sufficiently high for the interaction with the condensate to be
negligible. As we will show, the master equation for sympathetic
cooling can be written in terms of two processes: (a) Creation and
annihilation of particles in the ground state and (b) Creation and
annihilation of quasiparticles. From this master equation we
derive a rate equation for this number-conserving system under
sympathetic cooling which includes the growth, scattering and loss
(sympathetic evaporation) processes. This result is very
important, specially now that sympathetic cooling is used as an
alternative cooling method. In addition, this rate equation is
more general than the one obtained by Gardiner {\it et. al.} in
refs.\cite{qkt3,l1}, which is only valid in the regime of a highly
populated condensate \cite{newbo}. 

The paper is organized as follows: In section \ref{I} the
description of the sympathetic cooling model is given. In section
\ref{II} the master equation describing the dynamics of a
number-conserving system in the condensate band is derived. In
subsections \ref{growth}, \ref{scatt}, and \ref{loss} a rate equation
describing the population growth, the effects of the scattered
particles as well as the effects of the sympathetic evaporation
(trap losses) is derived. Section \ref{con} contains the
conclusion.

\section{Description of the system}
\label{I}

For the sake of completeness and notation we briefly review the
main results of Refs.~\cite{master,pap01}. However, we introduce
some modifications to the model presented in Ref.\cite{master}:
the cooling agent is a gas trapped in a harmonic potential, instead
of a free gas, and the gas to be cooled is confined in an open trap.
This situation is closer to the actual experimental conditions.

A system $A$ of $N_A$ bosons is subject to sympathetic cooling due
to its interaction with system $B$. The cooling agent of $N_B$
atoms is in thermal equilibrium at temperature $T_{B}$. The
single-particle states of system $B$ are then described by
harmonic oscillator wave functions with energy $\epsilon_{\vec
\ell}= \hbar\omega_{B}(\ell_{x}+\ell_{y}+\ell_{z})$, where
$\omega_{B}$ is the trap frequency and $\vec{\ell} =
(\ell_x,\ell_y,\ell_z)$ are the quantum numbers related to the
bath. The corresponding Hamiltonian can be written
\[H_B=\sum_{\vec{\ell}} \hbar \omega_{B} (\ell_x+\ell_y+\ell_z)
b^{\dagger}_{\vec{\ell}} b_{ \vec{\ell}},\] with the creation and
annihilation operators for the bath $b^{\dagger}_{\vec{\ell}}$ and
$b_{\vec{\ell}}$, respectively.

System $A$ is assumed to be confined in an open trap. For energies
$\epsilon$ much smaller than the trap depth $\epsilon_{t}$, the
trap can be approximated by a harmonic potential with trap
frequency $\omega_{A}$. The single-particle states in the
harmonic oscillator potential have the quantum numbers $\vec{n} =
(n_x,n_y,n_z)$. The single--particle eigenfunctions are labelled
$\psi_{\vec{n}}(\vec{x})$, the eigenvalues 
$\epsilon_{\vec n}=\hbar\omega_{A}
(n_x+n_y+n_z)$. The
associated creation and annihilation operators are denoted by
$a^{\dagger}_{\vec{n}}$ and $a_{\vec{n}}$, respectively. 
%The
%Hamiltonian has the form \[H_A=\sum_{\vec{n}} \hbar \omega_{A}
%(n_x+n_y+n_z) a^{\dagger}_{\vec{n}} a_{ \vec{n}} \quad
%\textrm{for}\quad \epsilon \ll \epsilon_{t}.\] 
If, on the other
hand, the energy is much larger than the trap depth, the system
$A$ can be approximately described by free particles . In this
case the states are described by plane waves with energy
$\epsilon({\vec \kappa})={\hbar^2}\vec{\kappa}^2/(2m)$, with
$\vec{\kappa}$ denoting the wavevector of the particle. 

\begin{equation}  \label{ha}
H_A=\sum_{\vec{n}} U_{\vec n} a^{\dagger}_{\vec{n}} a_{
\vec{n}}
\end{equation}

\noindent where

\begin{displaymath}
U_{\vec n}=\left\{ \begin{array}{ll}
\epsilon_{\vec n}& \textrm{if $\epsilon\ll\epsilon_{t}$} \\
\epsilon({\vec \kappa}) & \textrm{if $\epsilon\gg\epsilon_{t}$}
\end{array} \right.
\end{displaymath}
 
\noindent when $\epsilon\approx\epsilon_{t}$, $U_{\vec n}$ cannot be determined in a
simple way. 
%Now the
%Hamiltonian is
%\[H_A=\int d\vec{\kappa}\, \epsilon(\vec{\kappa})\, a^{\dagger}(\vec{\kappa}) %a(\vec{\kappa}) \quad \textrm{for}\quad
%\epsilon \gg \epsilon_{t}.\]

The two-body interaction between the bath and the particles can be
written in the form \[V_{A-B}=
\sum_{\vec{n},\vec{n'},\vec{\ell},\vec{\ell'}}\gamma_{\vec{n}
,\vec{n'},\vec{\ell},\vec{\ell'}} a^{\dagger}_{\vec{p}}
a_{\vec{p'}} b^{\dagger}_{\vec{\ell}} b_{\vec{\ell'}}.\]For
simplicity, we assume a point-like interaction characterized by a
scattering length $a$. We are interested in scattering processes
between the bath and tightly bound particles, and processes in
which low-energy particles of the system are scattered into the
continuum through the interaction with the bath. In the first
case, the matrix elements of the two-body interaction have the
form
\begin{equation}\gamma_{\vec{n},\vec{n'},\vec{\ell},\vec{\ell^{\prime}}}=\frac{4\pi{\hbar}^2a}{
2\mu_{AB}} \int d\vec{x} \psi^{*}_{\vec{n}}(\vec{x})
\psi_{\vec{n'}}(\vec{x})\psi^{*}_{\vec{\ell}}(\vec{x})
\psi_{\vec{\ell'}}(\vec{x}). \end{equation}

\noindent Here, $\mu_{AB}=m_A m_B / (m_A + m_B)$ denotes the
reduced mass. For scattering processes into the continuum, the
matrix element can be written as
\begin{equation}
\gamma_{\vec{n},\vec{\ell},\vec{\ell^{\prime}}}(\vec{\kappa})=\frac{4\pi{\hbar}^2a}{
2\mu_{AB}{(2\pi)}^{3/2}} \int d\vec{x} \quad e^{-i{\vec \kappa}\cdot{\vec
x}} \psi_{\vec{n}}(\vec{x })\psi^{*}_{\vec{\ell}}(\vec{x})
\psi_{\vec{\ell'}}(\vec{x}). \label{estrelinha}
\end{equation}

%Note that in the case of evaporative cooling the most important interaction is
%the one between the atoms in the same system $A$, i.e., $H_{A-A}$, so as a
%consequence the interaction above $H_{A-B}$ should be replace
%$H_{A-A}$. Actually, this is the case described in the series of papers using
%the Kinetic theory \cite{qkt3,l1,qkt6,l2,qkt1,qkt2,qkt4,qkt5,qkt7,l3,j1}.

As derived in Ref.~\cite{master}, the master equation for the
time dependence of the reduced density matrix $\rho_A(t)$,
obtained by tracing the total density matrix over the bath, has
the form

\begin{equation}
\label{eq1} \frac{\mathrm{d} \rho_A(t)}{\mathrm{d}t} = -
\frac{i}{\hbar} \biggl [ H_A + V^{'}_{A-A}, \rho_{A}(t) \biggr ] +
{\mathcal{L}} \rho_{A}(t),
\end{equation}

\noindent where $V^{'}_{A-A}$ describes the mutual interaction of the
atoms in system $A$. The interaction includes the two-body
interaction $V_{A-A}$ and, in addition, terms corresponding to
shifts produced by the elimination of the bath in the master
equation. It is assumed that the interaction $V_{A-A}$ is weak
compared to $V_{A-B}$, i.e. the thermalization of the system is
solely determined by the interaction with the bath. The equation
above is obtained under the the following assumptions: (i) Markov:
the correlation time for the
interaction between systems $A$ and $B$ is much shorter than the cooling
time, and (ii) rotating--wave approximation: 
terms rotating at multiples of the
trap frequency are neglected. 

It was shown in Ref.~\cite{pap01} that if the number $N_B$ of
particles in the bath is very large, decoherence acts very quickly
compared to the equilibration time and reduces the density matrix
to diagonal form. If only scattering processes involving tightly
bound atoms of system $A$ are involved, the Liouvillian in
Eq.~(\ref{eq1}) becomes

\begin{eqnarray}
\label{3} \mathcal{L} \rho_A & =& \sum_{{\vec m} \neq {\vec n}}
\Gamma^{{\vec m},{\vec n}} \biggl ( 2
a^{\dagger}_{\vec m} a_{{\vec n}} \rho_A(t) a^{\dagger}_{\vec n} a_{{\vec m}} - a^{\dagger}_{\vec n} a_{{\vec m%
}} a^{\dagger}_{\vec m} a_{{\vec n}} \rho_A(t)  \nonumber \\
& & \qquad \qquad \qquad \qquad \qquad - \rho_A(t) a^{\dagger}_{\vec n} a_{{%
\vec m}} a^{\dagger}_{\vec m} a_{{\vec n}} \biggr ), \ \label{eq3}
\end{eqnarray}

\noindent with the rate coefficients

\begin{eqnarray}
\label{rate} \Gamma^{{\vec m},{\vec n}}= \frac{1}{2 \hbar^2}
\int_{-\infty}^{\infty} {\rm d} \tau \sum_{\vec{\ell},\vec{\ell'}}
\ \gamma_{{\vec n},{\vec m},{\vec \ell},{\vec \ell'}}
 \ \gamma_{{\vec m},{\vec n},{\vec \ell'},{\vec \ell}} \nonumber
 \\
 \times \textsf{n}_{\vec{\ell}}[\textsf{n}_{\vec{\ell'}}+1]
\exp{\left[ i \left(\Delta\epsilon_{A}+ \Delta\epsilon_{B} \right)
\tau / \hbar
  \right]}
\end{eqnarray}

\noindent where $\textsf{n}_{\vec{\ell}}$ is the average
occupation number of the heat bath oscillator. The energy
difference for the system and the bath is given by
$\Delta\epsilon_{A}=\epsilon_{\vec{n}}-\epsilon_{\vec{m}}$ and
$\Delta\epsilon_{B}=\epsilon_{\vec{\ell}}-\epsilon_{\vec{\ell'}}$,
respectively. If particles are scattered into the continuum, the
rate coefficients have similar form but with the matrix element
defined in Eq.~(\ref{estrelinha}). More details about the evaluation
of the rate coefficients are given in Refs.\cite{master,pap01}.

\section{Master Equation for the Condensate Band}
\label{II}

\begin{figure}[h]
  \begin{center}
    \leavevmode
    \parbox{0.5\textwidth}
           {\psfig{file=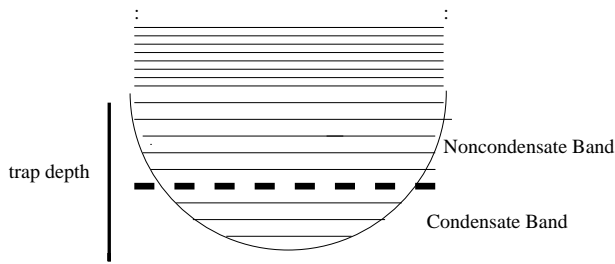}}
%width=0.4\textwidth,angle=270}}
  \end{center}
\protect\caption{Distribution of the trap levels in the condensate $B_{C}$
and noncondensate $B_{NC}$ bands.}
\label{fig1a}
\end{figure}

In this section, a master equation describing the dynamics of the
condensate band is derived. Basically, we extend the formalism
developed by Gardiner et al. \cite{qkt3} to the case of two
distinct gases. The trap-levels of the system $A$ are grouped in
two bands: the {\it condensate} band ($B_{C}$) and the band of
{\it noncondensed} particles ($B_{NC}$), where the latter is in
thermal equilibrium with the cooling agent system $B$ 
(see Figure \ref{fig1a}). 

The condensate band $B_{C}$ includes the ground state and all excited levels
in the trap which are directly influenced by the presence of the condensate. 
The noncondensate band $B_{NC}$ is composed by the highly excited levels,
which include also the continuum states for $\epsilon>\epsilon_{t}$.
The following assumptions are used in order to derive a master
equation for the condensate band from Eq.~(\ref{3}):

\begin{description}

\item (i) We neglect the correlations between the particles in the condensate
band and the particles in the noncondensate band. Therefore, we assume that
the complete density matrix of the system $A$ can be written as a direct
product between the density of the condensate band and the density
of the noncondensate band, i.e.,
$\rho_{A}=\rho_{C}\otimes\rho_{NC}$. Since the main interest is
in the dynamics of the condensate band, the noncondensate band is
eliminated from the description by tracing out the noncondensate
variable, i.e., $\rho_{C}=\mathrm{Tr}_{NC}(\rho_{A})$.

\item (ii) The density matrix of the noncondensate band $\rho_{NC}$
is considered in thermal equilibrium with the cooling agent at
temperature $T_{B}$. The levels of this band thermalize much
faster with the cooling agent than the levels of the condensate
band. Due to this assumption, we will completely neglect
collisions between particles of the noncondensate band and
particles of the cooling agent.

\end{description}

\begin{figure}[h]
  \begin{center}
    \leavevmode
    \parbox{0.5\textwidth}
           {\psfig{file=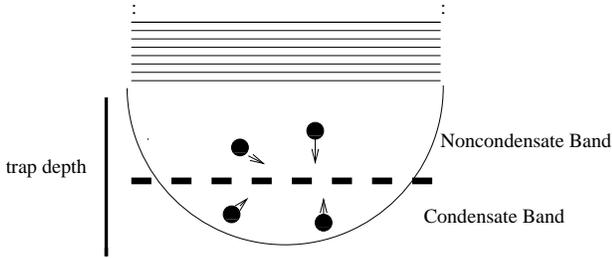}}
%width=0.4\textwidth,angle=270}}
  \end{center}
\protect\caption{Collisions between
particles in the condensate band $B_{C}$ and in the noncondensate band
$B_{NC}$}
\label{fig2}
\end{figure}

\begin{figure}[h]
  \begin{center}
    \leavevmode
    \parbox{0.5\textwidth}
           {\psfig{file=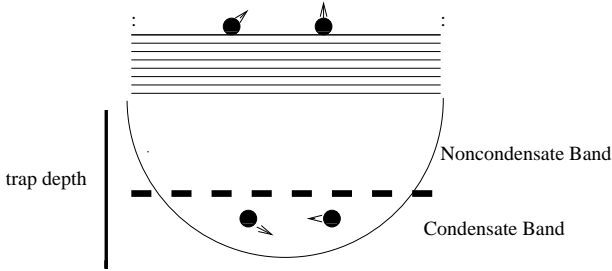}}
%width=0.4\textwidth,angle=270}}
  \end{center}
\protect\caption{Loss of particles from the trap, due to 
their interaction with the cooling agent atoms.}
\label{fig3}
\end{figure}

\begin{figure}[h]
  \begin{center}
    \leavevmode
    \parbox{0.5\textwidth}
           {\psfig{file=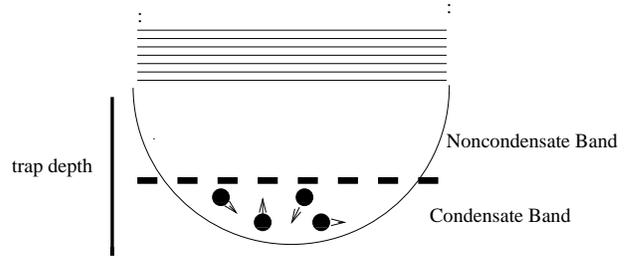}}
%width=0.4\textwidth,angle=270}}
  \end{center}
\protect\caption{Collisions between particles in the condensate band $B_{C}$.}
\label{fig4}
\end{figure}

\noindent Applying these assumptions to Eq.(\ref{3}), a master
equation for the condensate band can be derived,

\begin{eqnarray}
\label{eq4} \frac{d\rho_{C}(t)}{dt}&=& \sum_{\begin{array}{c}
{\vec m}\in {\rm B_{NC}}\\
{\vec n}\in {\rm B_{C}}\end{array}} \Gamma^{{\vec{n}},{\vec m}} \
{\langle N_{\vec m}\rangle}_{\textsf{n}_{0}} \biggl(
2 a^{\dagger}_{\vec n} \rho_C(t) a_{\vec n} \nonumber \\
&&-a_{n}a^{\dagger}_{n} \rho_C(t)
-\rho_C(t)a_{\vec n}a^{\dagger}_{\vec n} \biggr) \nonumber \\ \label{eq43} \\
&& + \sum_{\begin{array}{c}
{\vec m} \in {\rm B_{NC}}\\
{\vec n} \in {\rm B_{C}} \end{array}} \Gamma^{{\vec m},{
\vec{n}}}\ {\langle N_{\vec m}+1\rangle}_{\textsf{n}_{ 0}}
\biggl( 2 a_{\vec n}\rho_C(t) a^{\dagger}_{\vec n}\nonumber \\
&&-a^{\dagger}_{\vec n}a_{\vec n} \rho_C(t)
-\rho_C(t)a^{\dagger}_{\vec n}a_{\vec n} \biggr)\nonumber \\
\label{eq44} \\ \nonumber &&
+ \sum_{\begin{array}{c} {\vec m} \neq
{\vec n}\\{\vec n},{\vec m} \in {\rm B_{C}} \end{array} }
\Gamma^{{\vec m},{\vec n}} \ \biggl ( 2 a^{\dagger}_{\vec m}
a_{{\vec n}}
\rho_C(t) a^{\dagger}_{\vec n} a_{{\vec m}} \nonumber\\
&&- a^{\dagger}_{\vec n} a_{{\vec m}} a^{\dagger}_{\vec m} a_{{\vec n}} \rho_C(t)
- \rho_C(t) a^{\dagger}_{\vec n} a_{{%
\vec m}} a^{\dagger}_{\vec m} a_{{\vec n}} \biggr ) \nonumber \\
\label{eq45} \\ \nonumber
&&+ \sum_{\begin{array}{c}{\vec m} \neq {\vec n}\\
{\vec n},{\vec m} \in {\rm B_{C}} \end{array}} \Gamma^{{\vec
n},{\vec m}} \ \biggl ( 2 a^{\dagger}_{\vec n} a_{{\vec m}}
\rho_C(t) a^{\dagger}_{\vec m} a_{{\vec n}}\nonumber \\
&&- a^{\dagger}_{\vec m} a_{{\vec n }} a^{\dagger}_{\vec n}
a_{{\vec m}} \rho_C(t) - \rho_C(t) a^{\dagger}_{\vec m} a_{{ \vec
n}} a^{\dagger}_{\vec n} a_{{\vec m}} \label{eq46}\biggr )
\end{eqnarray}

\noindent where $\Gamma^{{\vec{m}},{\vec n}}=\Gamma^{{\vec{n}},{\vec
m}}e^{-\beta(\epsilon_{\vec m}-\epsilon_{\vec n})}$ and 
${\langle N_{\vec m}\rangle}_{\textsf{n}_{0}}\equiv \mathrm{Tr} 
(a_{\vec{m}}^\dagger a_{\vec{m}} \rho_{NC})_{\textsf{n}_{0}}$ is 
the expectation value
of the particle number operator for state $\vec m$, given that
there are $\textsf{n}_{0}$ bosons in the ground state. Since the
terms with $\vec m=\vec n$ cancel, this case is already excluded
from the summation. The first two terms,
Eqs.(\ref{eq43},\ref{eq44}), account for collisions between
particles in the condensate band and in the noncondensate band,
and are illustrated in Fig.\ref{fig2}. The loss of particles
from the trap, due to their interaction with the cooling agent atoms
are also included in Eq.(\ref{eq44}). The existence of 
such a process is related to the fact that the
trap is open, hence the atoms with higher energy than the trap energy 
can indeed escape. In the present model these particles are 
considered to escape to continuum states (non-bound states). In this case
the single-particle states are described by plane waves. Evaporation is
illustrated in Fig.\ref{fig3}.
The following two terms,
Eqs.(\ref{eq45},\ref{eq46}), describe collisions only between
particles in the condensate band, which are shown in Fig.\ref{fig4}.

%They represent the scattering
%process between atoms that are in the levels of the condensate
%band due its interaction with the atoms of the cooling agent
%(bath). Those terms also give important contribution during the
%process of formation of the condensate. Note that the scattering
%between atoms in degenerate energy levels $\vec m=\vec n$ will not
%give any contribution. Since the terms with $\vec m=\vec n$
%cancel, the case $\vec m =\vec n$ is already excluded from the
%summation. For Eqs.(\ref{eq4}-\ref{eq46}) the rate coefficients
%are defined by Eq.(\ref{ratemenor}).
%are defined by Eq.(\ref{ratemenor}). Finally, the last term
%represents the trap losses and is obtained from the assumption
%described in (C) c-), where ${\vec m}$ describes the untrapped
%states (continuum). In this case the rate coefficients are
%described by Eq.(\ref{ratemaior}). In summary,

Eqs.~(\ref{eq4}-\ref{eq46}) give an accurate treatment of the
internal dynamics of the condensate band, describing how the
condensate forms and how the particles are scattered by
collisions. Effects of elementary excitations of the condensate,
representing the ``thermal cloud'', do not appear explicitly,
since they are hidden in the single-particle operators
$a_{\vec n}$ and $a^{\dagger}_{\vec n}$ for $\vec n \in
\mathrm{B_C}$.
%In other words, those effects are intrinsic in the operators
%$a_{\vec n}$ and $a^{\dagger}_{\vec n}$ due to their relation with
%$\hat{\beta}_{\vec n}$ and $\hat{\beta_{\vec n}^{\dagger}}$.
In the limit of a highly condensed ground state, i.e.
${\mathsf{n_0}} \simeq N$, the effects of the condensate
excitations can be totally neglected \cite{l1}. During the process
of condensate formation, however, their effects play a crucial
role.

To analyse the role of elementary excitations, we resort to a
formalism of Girardeau and Arnowitt~\cite{gira}, which conserves
the number of particles. The method is based on the annihilation
$\hat{\beta}_0$ and creation $\hat{\beta}_0^{\dagger}$ operators
of one particle in the ground state, defined as $\hat{\beta}_0=
{(\hat{N}_0+1)}^{-1/2}\hat{a}_{0}$  and  $\hat{\beta}_0^{\dagger}=
\hat{a}_{0}^{\dagger}{(\hat{N}_0+1)}^{-1/2}$, and in the phonon
operators $\hat{\beta}_{\vec n}$, $\hat{\beta}^{\dagger}_{\vec n}$
defined as $\hat{\beta}_{\vec
n}=\hat{\beta}_0^{\dagger}\hat{a}_{\vec n}$ and $\hat{\beta}_{\vec
n}^{\dagger}= \hat{a}_{\vec n}^{\dagger}\hat{\beta}_0$, where
$\hat{N}_{0}$ is the number operator of particles in the ground-state,
$a_{\vec{n}}$ and $a^{\dagger}_{\vec{n}}$ are usual annihilation
and creation operators for the trap level $\vec n$. These operators 
obey the Bose commutation relations,
$[\hat{\beta}_{\vec n},\hat{\beta}_{\vec
n'}^{\dagger}]=\delta_{\vec{n},\vec{n'}}$. The phonons operators
can be written in terms of the creation and annihilation
quasiparticles operators: $\hat{b}_{\vec n}$ and
$\hat{b}^{\dagger}_{\vec n}$ , i.e., $\hat{\beta}_{\vec n}=u_{\vec
n}\hat{b}_{\vec n}+v_{\vec n} \hat{b}^{\dagger}_{\vec n}$ and
$\hat{\beta}^{\dagger}_{\vec n}=u^{*}_{\vec
n}\hat{b}^{\dagger}_{\vec n} +v^{*}_{\vec n}\hat{b}_{\vec n}$,
with ${|u_{\vec n}|}^2 -{|v_{\vec n}|}^2=1$. The formalism of
Girardeau and Arnowitt is more general than the one developed by
Gardiner \cite{newbo}, since it covers all cases ranging from a
total absence of population in the ground state $\textsf{n}_{0}=0$
to a highly populated ground state $\textsf{n}_{0}\simeq N$. A
detailed comparison of both formalisms can be found in
Ref.\cite{gira1}.

We replace the operators
$\hat{a}_0,\hat{a}_0^{\dagger},\hat{a}_{\vec n},\hat{a}_{\vec
n}^{\dagger}$ in Eqs.~(\ref{eq4}-\ref{eq46}) by the
particle-number-conserving Girardeau-Arnowitt operators
$\hat{\beta}_0,\hat{\beta}_0^{\dagger},\hat{\beta}_{\vec
n},\hat{\beta}_{\vec n}^{\dagger}$. Then, the action of the
opertators on the density matrix $\rho_{C}(t)$ is computed. Since
the decoherence time is very fast (see discussion in
Ref.~\cite{pap01}), only the diagonal terms of the reduced density
matrix $\rho_C$ contribute to the dynamics. Hence, only diagonal
terms will be considered. In this way, we obtain a master equation
for the diagonal elements $P(\textsf{n}_{0},\vec{n})=\langle
\textsf{n}_{0},\vec{n}|\rho_{C}(t)|\textsf{n}_{0},\vec{n}\rangle$,
with
$\vec{n}\equiv(\textsf{n}_{1},...,\textsf{n}_{m},\textsf{n}_{n},...)$
denoting the set of occupation number for all quasiparticles:

\begin{eqnarray}
\label{dia2}
\frac{d}{dt}P(\textsf{n}_{0},\vec{n},t)&=&\frac{d}{dt}P_{\rm
growth}(\textsf{n}_{0},\vec{n},t) +\frac{d}{dt}P_{\rm
scatt}(\textsf{n}_{0},\vec{n},t)\nonumber\\&&+\frac{d}{dt}P_{\rm
evap}(\textsf{n}_{0},\vec{n},t).
\end{eqnarray}

\noindent The rate $dP_{\rm growth}(\textsf{n}_{0},\vec{n},t)/dt$
represents the contribution of the condensate itself and the
quasiparticles for the growth of the condensate. The number of
particles in the condensate band changes. $dP_{\rm
scatt}(\textsf{n}_{0},\vec{n},t)/dt$ describes scattered particles
in the condensate band. During the scattering process the number
of particles in the condensate band does not change. Finally,
$dP_{\rm evap}((\textsf{n}_{0},\vec{n},t)/dt$ includes the
evaporation of particles with energies larger than the trap depth.
%In the next subsection, it will be clear the form of such
%terms, since this master equation for the occupation probability is discussed.%In
%the first subsection, the contribution of the first two terms of the master
%equation, namely, Eqs.(\ref{eq43}$-$\ref{eq44}) are analized.
%They give information about the growth process. In the second subsection, the
%contribution of the
%scattered terms, namely, Eqs.(\ref{eq45}$-$\ref{eq46}) are given. In the third
%subsection, the effects of the evaporation Eq.(\ref{eq47}) are taken
%into account. These
%effects together give information about the rate equation for the population
%of the system.

\section{Rate Equations for the Condensate Band}
\label{IIa}

\subsection{Growth} \label{growth}

Consider the diagonal elements of the 
Eqs.~(\ref{eq43}$-$\ref{eq44}) for $\vec{n}=0$ and $\vec{n}\neq
0$ with the replacements described in the previous section, we arrive
at a master equation for the occupation probability of the
condensate band

\begin{equation}
\label{pgrow} \frac{d}{dt}P_{\rm
growth}(\textsf{n}_{0},\vec{n},t)=R_{C}+R_{Q},
\end{equation}

\noindent where $R_{C}$ represents the growth rate of the
condensate itself

\begin{eqnarray}
\label{a0}
R_{C}&&=2\textsf{n}_{0}\Gamma^{+}(\textsf{n}_{0}-1)P(\textsf{n}_{0}-1,\vec{n})
-2(\textsf{n}_{0}+1)\Gamma^{+}(\textsf{n}_{0})P(\textsf{n}_{0},\vec{n}) \nonumber \\
&&+
2(\textsf{n}_{0}+1)\Gamma^{-}(\textsf{n}_{0}+1)P(\textsf{n}_{0}+1,\vec{n})
-2\textsf{n}_{0}\Gamma^{-}(\textsf{n}_{0})P(\textsf{n}_{0},\vec{n}).
\nonumber\\
\end{eqnarray}

\noindent The coefficients $\Gamma^{+}(\textsf{n}_{0})$ and
$\Gamma^{-}(\textsf{n}_{0})$ carry information about particles
entering and leaving the condensate

\begin{eqnarray}
&&\Gamma^{+}(\textsf{n}_{0})\equiv\sum_{\vec{m}\neq
0}\Gamma^{{\vec 0},{\vec m}}{\langle N_{\vec m}\rangle}_{\textsf{n}_{0}}\nonumber, \\
&&\Gamma^{-}(\textsf{n}_{0})\equiv\sum_{\vec{m}\neq
0}\Gamma^{{\vec m},{\vec 0}}{\langle N_{\vec m}+1
\rangle}_{\textsf{n}_{0}}.\nonumber
\end{eqnarray}

\noindent The rate $R_{Q}$ takes into account the contribution of
the elementary excitations

\begin{eqnarray}
\label{ak} R_{Q}&&=2\sum_{\vec n} \textsf{n}_{\vec
n}\Gamma^{++}_{\vec n}(\textsf{n}_{0}-1,\vec{n}-\vec{e}_{\vec
n})P(\textsf{n}_{0}-1,\vec{n}-\vec{e}_{\vec n}) \nonumber
\\&&-2\sum_{\vec n}(\textsf{n}_{\vec n}+1)
\Gamma^{++}_{\vec n}(\textsf{n}_{0},\vec{n})P(\textsf{n}_{0},\vec{n})\nonumber \\
&&+2\sum_{\vec n} (\textsf{n}_{\vec n}+1)\Gamma_{\vec
n}^{--}(\textsf{n}_{0}+1,\vec{n}+\vec{e_{\vec
n}})P(\textsf{n}_{0}+1,\vec{n}+\vec{e_{\vec n}})\nonumber
\\&&-2\sum_{\vec n}\textsf{n}_{\vec n}\Gamma_{\vec n}^{--}P(\textsf{n}_{0},\vec{n})
\nonumber \\
&&+2\sum_{\vec n} (\textsf{n}_{\vec n}+1)\Gamma^{+-}_{\vec
n}(\textsf{n}_{0}-1,\vec{n}+\vec{e_{\vec
n}})P(\textsf{n}_{0}-1,\vec{n}+\vec{e_{\vec n}})\nonumber \\&&
-2\sum_{\vec n}\textsf{n}_{\vec n}\Gamma^{+-}_{\vec
n}(\textsf{n}_{0})P(\textsf{n}_{0},\vec{n})
\nonumber \\
&&+2\sum_{\vec n}\textsf{n}_{\vec
n}\Gamma^{-+}_{n}(\textsf{n}_{0}+1,\vec{n}-\vec{e_{\vec
n}})P(\textsf{n}_{0}+1,\vec{n}-\vec{e_{\vec n}})\nonumber \\&&
-2\sum_{\vec n}(\textsf{n}_{\vec n}+1) \Gamma^{-+}_{\vec
n}(\textsf{n}_{0},\vec{n})P(\textsf{n}_{0},\vec{n}),
\end{eqnarray}

\noindent with $\vec{e_{n}}\equiv(...,0,0,1,0,0,...)$. The
coefficients $\Gamma^{++}_{\vec n}(\textsf{n}_{0},\vec{n})$,
$\Gamma^{--}_{\vec n}(\textsf{n}_{0},\vec{n})$, $\Gamma^{+-}_{\vec
n}(\textsf{n}_{0},\vec{n})$ and $\Gamma^{-+}_{\vec
n}(\textsf{n}_{0},\vec{n})$ are defined as

\begin{eqnarray}
&&\Gamma^{++}_{\vec
n}(\textsf{n}_{0},\vec{n})\equiv\sum_{\vec{m}\neq
\vec{n}}\Gamma^{{\vec n},{\vec m}}{\langle N_{\vec m}\rangle}_{\textsf{n}_{0}}{|u_{\vec n}|}^2, \nonumber \\
&&\Gamma^{--}_{\vec
n}(\textsf{n}_{0},\vec{n})\equiv\sum_{\vec{m}\neq
\vec{n}}\Gamma^{{\vec m},{\vec n}}
{\langle N_{\vec m}+1\rangle}_{\textsf{n}_{0}} {|u_{\vec n}|}^2, \nonumber \\
&&\Gamma^{+-}_{\vec
n}(\textsf{n}_{0},\vec{n})\equiv\sum_{\vec{m}\neq
\vec{n}}\Gamma^{{\vec n},{\vec m}}{\langle
N_{\vec m}\rangle}_{\textsf{n}_{0}} {|v_{\vec n}|}^2, \nonumber \\
&&\Gamma^{-+}_{\vec
n}(\textsf{n}_{0},\vec{n})\equiv\sum_{\vec{m}\neq
\vec{n}}\Gamma^{{\vec m},{\vec n}} {\langle N_{\vec
m}+1\rangle}_{\textsf{n}_{0}} {|v_{\vec n}|}^2,\nonumber
\end{eqnarray}

\noindent where $\vec m \in B_{NC}$ and  $\vec n \in B_{C}$. These 
coefficients describe processes of creation
and annihilation of phonons inside the condensate.

In order to obtain the dynamics of the population in the
condensate band, the mean number of particles is evaluated. We
will start with the contribution of the condensate mode itself. In
Eq.(\ref{a0}) we use the thermodynamic relation between the rate
coefficients

\[
\Gamma^{+}(\textsf{n}_{0})=\Gamma^{-}(\textsf{n}_{0})e^{\beta(\epsilon_{0}-\mu_{0})}
\]

\noindent in combination with the factorization assumption
$P(\textsf{n}_{0},\vec{n})=P(\textsf{n}_{0})\otimes
P(\vec{n})=P(\textsf{n}_{0})\ldots\otimes P(\textsf{n}_{n})\otimes
\ldots$. Here $\beta=1/k_{B}T_{B}$, $\epsilon_{0}$ is the energy
of the ground state and $\mu_0 \equiv \mu(\mathsf{n_0})$ the
chemical potential for $\mathsf{n_0}$ particles in the ground
state. For the rate of change of the condensate number one then
finds

\begin{eqnarray}
\label{n} \frac{d\textsf{n}_{0}}{dt}|_{\rm growth}&&=2\biggl(
(\textsf{n}_{0}+1)\Gamma^{+}(\textsf{n}_{0})-\textsf{n}_{0}\Gamma^{-}(\textsf{n}_{0})\biggr)\nonumber \\
&&=2\Gamma^{+}(\textsf{n}_{0})\biggl(\textsf{n}_{0}(1-e^{-\beta(\mu_{0}-\epsilon_{0})})
+1\biggr).
\end{eqnarray}

\noindent In the limit of ${\mathsf{n}_{0}}\simeq N$, with $N \gg 1$,
the equation above can be approximated by

\begin{equation}
\label{noap} \frac{dN}{dt}\approx 2\Gamma^{+}(N)N
\biggl((1-e^{\beta(\mu_0-\epsilon_{0})})\biggr),
\end{equation}

\noindent where the coefficient $\Gamma^{+}(N)$ contains
information about the sympathetic cooling process between system
$A$ and cooling agent $B$.

If we apply the formalism described here to a gas of only one
species, we find a rate equation for the growth of the condensate
which has a form similar to the one obtained by Gardiner and
coworkers based on \cite{newbo} (see Eq.~(23) in Ref.~\cite{l1}
and Eq.(168) in Ref.\cite{qkt3}). However, due to the properties of the
Giradeau-Arnowitt formalism, the rate equation in our case is not
limited only to highly populated condensates but also includes
condensates of arbitrary particle number.

%If only collisions between particles of the same species were
%taken into account, one would recover the result of Gardiner and
%co-workers (see Eq.~(23) in Ref.~\cite{l1} and Eq.(168) in
%Ref.\cite{qkt3}) who, instead of the Giradeau-Arnowitt formalism,
%used the Gardiner's approach \cite{newbo}.

We now explore the contribution of the elementary excitations for the growth
process. We use the thermodynamic relation
between the coefficients

\[
\Gamma^{++}_{\vec n}(\textsf{n}_{0},\vec{n})=\Gamma^{--}_{\vec
n}(\textsf{n}_{0},\vec{n})e^{\beta(\epsilon_{\vec n}-\mu_{0})},
\]

\[
\Gamma^{+-}_{\vec n}(\textsf{n}_{0},\vec{n})=\Gamma^{-+}_{\vec
n}(\textsf{n}_{0},\vec{n})e^{\beta(\epsilon_{\vec n}-\mu_{0})},
\]

\noindent and factorize all correlations i.e.,
$P(\textsf{n}_{0},\vec{n})=P(\textsf{n}_{0})\otimes
...P(\textsf{n}_{n})...$ in Eq.(\ref{ak}). Thus, the rate equation
for the elementary excitations is given by

\begin{eqnarray}
\label{taxaQ} \frac{d\textsf{n}_{\vec{n}}}{dt}|_{\rm
growth}&&=2\sum_{\vec{n}}\Gamma^{++}_{\vec{n}}(\textsf{n}_{0},\vec{n})\biggl((1-e^{\beta(\mu_{0}-\epsilon_{\vec
n})})\textsf{n}_{\vec n}+1\biggr)
\nonumber \\
&&-2\sum_{\vec n}\Gamma^{+-}_{\vec
n}(\textsf{n}_{0},\vec{n})\biggl((1-e^{\beta(\mu_{0}-\epsilon_{\vec
n})})\textsf{n}_{\vec n}+1\biggr).
\end{eqnarray}

\noindent A simplified form of Eq.(\ref{taxaQ}) can be derived by
using the relation
$\Gamma^{+-}_{\vec{n}}(\textsf{n}_{0},\vec{n})=\Gamma^{++}_{\vec{n}}(\textsf{n}_{0},\vec{n})-\Gamma^{+}_{\vec{n}}(\textsf{n}_{0},\vec{n})$,

\begin{eqnarray}
\label{taxaQsim} \frac{d\textsf{n}_{\vec{n}}}{dt}|_{\rm
growth}&&=2\sum_{\vec{n}}\Gamma^{+}_{\vec{n}}
(\textsf{n}_{0},\vec{n})\biggl((1-e^{\beta(\mu_{0}-\epsilon_{\vec
n})})\textsf{n}_{\vec n} +1\biggr),
\end{eqnarray}

\noindent with $\Gamma^{+}_{\vec{n}}(\textsf{n}_{0},\vec{n})=\sum_{\vec{m}}
\Gamma^{\vec n, \vec m}{\langle N_{\vec m}\rangle}_{\textsf{n}_{0}}$.\\

The total growth equation can be obtained from the contribution of
the condensate mode itself Eq.(\ref{n}) and the elementary
excitations Eq.(\ref{taxaQsim}).

\subsection{Scattering}
\label{scatt}

In this subsection the effects of the scattering processes of
atoms inside the condensate band are considered, as described by
Eqs.~(\ref{eq45}$-$\ref{eq46}). During the scattering process the
number of particles in the condensate band does not change.

We use the same analytical procedure described in the previous
section to derive a rate equation related to the scattering
process. However, we will omit details of this calculation, since the 
expressions can be rather large. 
%Details of this calculation can be found in Appendix \ref{a}.
Thus, the corresponding scattering rate equation for
$\textsf{n}_{\vec n}=\langle a_{\vec n}^{\dagger}a_{\vec
n}\rangle$ has the form

\begin{eqnarray}
\label{nscatt} \frac{d\textsf{n}_{\vec
n}}{dt}|_{\rm{scatt}}&&=2\sum_{\vec{n}\neq 0}\Gamma_{\vec n}^{u+}
\textsf{n}_{0}[e^{-\beta(\epsilon_{\vec n}-\epsilon_{0})})-\textsf{n}_{\vec n}
(1-e^{-\beta(\epsilon_{\vec n}-\epsilon_{0})})]\nonumber \\
&&+2\sum_{\vec{n}\neq 0}\Gamma_{\vec n}^{v+}
\textsf{n}_{0}[\textsf{n}_{\vec n}(1-e^{-\beta(\epsilon_{\vec n}-\epsilon_{0})})+1]\nonumber \\
&&+2\sum_{\vec{m}\neq \vec{n}\neq 0}\Gamma_{{\vec m},{\vec
n}}^{uu+-}[\textsf{n}_{\vec m}(\textsf{n}_{\vec n}+1)\nonumber
\\&&-\textsf{n}_{\vec n}(\textsf{n}_{\vec
m}+1)e^{-\beta(\epsilon_{\vec m}-\epsilon_{\vec n})}]\nonumber\\
&&+2\sum_{\vec{m}\neq \vec{n}\neq0}\Gamma_{{\vec m},{\vec
n}}^{vv+-}[-\textsf{n}_{\vec n}(\textsf{n}_{\vec
m}+1)\nonumber \\&&+\textsf{n}_{\vec m}(\textsf{n}_{\vec n}+1)e^{-\beta(\epsilon_{\vec m}-\epsilon_{\vec n})}]\nonumber\\
&&+2\sum_{\vec{m}\neq \vec{n}\neq 0}\Gamma_{{\vec m},{\vec
n}}^{uv++}[(\textsf{n}_{\vec m}+1)(\textsf{n}_{\vec n}+1)\nonumber
\\&&-\textsf{n}_{\vec n}\textsf{n}_{\vec
m}e^{-\beta(\epsilon_{\vec m}-\epsilon_{\vec n})}]\nonumber\\
&&+2\sum_{\vec{m}\neq \vec{n}\neq 0}\Gamma_{{\vec m},{\vec
n}}^{vu++}[-\textsf{n}_{\vec n}\textsf{n}_{\vec m}\nonumber \\&&
+(\textsf{n}_{\vec m}+1)(\textsf{n}_{\vec
n}+1)e^{-\beta(\epsilon_{\vec m}-\epsilon_{\vec n})}],
\end{eqnarray}

\noindent with

\begin{eqnarray}
\label{coefscat} &&\Gamma_{\vec
n}^{u+}(\textsf{n}_{0},\vec{n})\equiv\Gamma^{{0},{\vec n}}_{{
\vec n},{0}}{|u_{\vec n}|}^2,\nonumber \\
&&\Gamma^{v+}_{\vec
n}(\textsf{n}_{0},\vec{n})\equiv\Gamma^{{0},{\vec
n}}_{{\vec n,0}}{|v_{\vec n}|}^2,\nonumber \\
&&\Gamma_{{\vec n},{\vec
m}}^{uu+-}(\textsf{n}_{0},\vec{n})\equiv\Gamma^{{\vec n},{\vec m}}{|u_{\vec n}|}^2{|u_{\vec m}|}^2,\nonumber\\
&&\Gamma_{{\vec n},{\vec
m}}^{vv+-}(\textsf{n}_{0},\vec{n})\equiv\Gamma^{{\vec n},{\vec m}}{|v_{\vec n}|}^2{|v_{\vec m}|}^2,\nonumber\\
&&\Gamma_{{\vec n},{\vec
m}}^{uv++}(\textsf{n}_{0},\vec{n})\equiv\Gamma^{{\vec n},{\vec m}}{|u_{\vec n}|}^2{|v_{\vec m}|}^2,\nonumber\\
&&\Gamma_{{\vec n},{\vec
m}}^{vu++}(\textsf{n}_{0},\vec{n})\equiv\Gamma^{{\vec n},{\vec m}}{|v_{\vec n}|}^2{|u_{\vec m}|}^2.
\end{eqnarray}

\noindent The first two terms of Eq.(\ref{nscatt}) account for the
scattering between the particles in the ground state and the
quasiparticles of the $\vec n$-level. The other terms describe the
scattering between quasiparticles of different levels.

%\noindent The following relations should be used to simplify Eq(\ref{nscatt})

%\begin{eqnarray}
%&&\Gamma^{vv+-}(\textsf{n}_{0},\vec{n})=\Gamma^{uu+-}(\textsf{n}_{0},\vec{n})-\Gamma_{1}-\Gamma_%{2}+\Gamma_{3}\nonumber \\
%&&\Gamma^{uv+-}(\textsf{n}_{0},\vec{n})=\Gamma^{uu+-}_{\vec{m},\vec{n}}(\textsf{n}_{0},\vec{n})-%\Gamma_{1}\nonumber \\
%&&\Gamma^{vu+-}(\textsf{n}_{0},\vec{n})=\Gamma^{uu+-}_{\vec{m},\vec{n}}(\textsf{n}_{0},\vec{n})-%\Gamma_{2}\nonumber\\
%\end{eqnarray}

%\noindent with

%\begin{eqnarray}
%&&\Gamma_{1}=\sum_{\vec n\neq \vec m}\Gamma_{\vec n, \vec m}^{\vec m, \vec n}{%|u_{n}|}^2 \nonumber \\
%&&\Gamma_{2}=\sum_{\vec n\neq \vec m}\Gamma_{\vec n, \vec m}^{\vec m, \vec n}{%|u_{m}|}^2 \nonumber \\
%&&\Gamma_{3}=\sum_{\vec n\neq \vec m}\Gamma_{\vec n, \vec m}^{\vec m, \vec n}
%\end{eqnarray}

\subsection{Evaporation}
\label{loss}

We now discuss the evaporation of particles from the trap. Atoms can
escape from the condensate band of the trap to unbound states.
This loss of particles from the trap is induced by their
interaction with the cooling agent. This process is particularly
important in the initial stage of the cooling process
\cite{allard,matthias}. To implement evaporation in the present
model, particles are assumed to escape to continuum states
(unbound states) inside the noncondensate band. Only
Eq.~(\ref{eq44}) contributes to this process, since the rate
coefficients of Eq.~(\ref{eq4}), which includes the overlap of the
wave functions involved in the process, is completely negligible.
Taking the diagonal elements of Eq.~(\ref{eq44}), $P_{\rm{evap}}$
is obtained

\begin{eqnarray}
\label{evap}
&&\frac{d}{dt}P_{\rm{evap}}(\textsf{n}_{0},\vec{n},t)=
-2\textsf{n}_{0}\gamma^{-}(\textsf{n}_{0})P(\textsf{n}_{0},\vec{n})\nonumber\\
&&+2(\textsf{n}_{0}+1)
\gamma^{-}(\textsf{n}_{0}+1)P(\textsf{n}_{0}+1,\vec{n})
\nonumber\\
&&+2\sum_{\begin{array}{c} {\vec n} \in {\rm
B_{C}}\end{array}}(\textsf{n}_{\vec
n}+1)\gamma^{--}_{\vec{n}}(\textsf{n}_{0}+1,\vec{n}+\vec{e}_{n})P(\textsf{n}_{0}+1,\vec{n}
+\vec{e}_{n})\nonumber\\
&&-2\sum_{\begin{array}{c} {\vec n} \in {\rm
B_{C}}\end{array}}\textsf{n}_{\vec
n}\gamma^{--}(\textsf{n}_{0})P(\textsf{n}_{0},\vec{n})\nonumber
\\&&+2\sum_{\begin{array}{c} {\vec n} \in {\rm
B_{C}}\end{array}}\textsf{n}_{\vec
n}\gamma^{-+}_{\vec{n}}(\textsf{n}_{0}+1,\vec{n}-\vec{e}_{n})
P(\textsf{n}_{0}+1,\vec{n}-\vec{e}_{n})\nonumber\\
&&-2\sum_{\begin{array}{c} {\vec n} \in {\rm
B_{C}}\end{array}}(\textsf{n}_{\vec n}+1)
\gamma^{-+}(\textsf{n}_{0},\vec{n})P(\textsf{n}_{0},\vec{n}),
\end{eqnarray}

\noindent where

\begin{eqnarray}
&&\gamma^{-}(\textsf{n}_{0},\vec{n})\equiv\sum_{ {\vec m} \in {\rm
B^{(t)}_{NC}}}
\Gamma^{\vec{m},0},\nonumber\\
&&\gamma^{--}_{\vec n}(\textsf{n}_{0},\vec{n})\equiv\sum_{{\vec m}
\in {\rm B^{(t)}_{NC}}} \Gamma^{{\vec m},{\vec n}}
{|u_{\vec n}|}^2,\nonumber\\
&&\gamma^{-+}_{\vec n}(\textsf{n}_{0},\vec{n})\equiv\sum_{{\vec m}
\in {\rm B^{(t)}_{NC}}}\Gamma^{{\vec m},{\vec n}}{|v_{\vec n}|}^2\nonumber\\
&&\quad\quad\quad\quad\quad=\gamma^{-+}_{\vec
n}(\textsf{n}_{0},\vec{n})-\gamma_{\vec n}^{-},
\end{eqnarray}

\noindent with $\rm B^{(t)}_{NC}$ denoting the noncondensate band
with energy $\epsilon_{\vec m}>\epsilon_{t}$. As mentioned at the
end of section \ref{I}, these rate coefficients $\Gamma^{{\vec
m},{0}}$ and $\Gamma^{{\vec m},{\vec n}}$ have a different form from
the ones which appear in the growth and scattering processes,
since now $\vec m$ describes unbound states. The first line of
Eq.(\ref{evap}) describes the evaporation of an atom out of the
ground state while the others describe the evaporation of an atom
out of the $n$-excited level of the condensate band.
%with $\vec{n} \in {\rm CB}$where the $m$-level represents the
%continuum state (untrapped state), as a consequence it is
%described by a plane wave. In this case, the $\Gamma^{{\vec
%m},{\vec n}}$ coefficients are given by Eq.(\ref{ratemaior}). The

Following the same procedure described in the previous
subsections, a rate equation for the evaporation of particles from
the ground state and the evaporation of quasiparticles in the
condensate band is obtained:

\begin{eqnarray}
\label{neva0}
&&\frac{d{\textsf{n}_{0}}}{dt}|_{\rm evap}=-2\quad\gamma^{-}(\textsf{n}_{0},\vec{n})\quad{\textsf{n}_{0}|}_{\rm evap},\nonumber\\
&&\frac{d{\textsf{n}_{\vec n}}}{dt}|_{\rm evap}=-2\sum_{\vec
n}\gamma^{-}_{\vec
n}(\textsf{n}_{0},\vec{n})\quad{\textsf{n}_{\vec n}|}_{\rm evap},
\end{eqnarray}

\noindent with

\begin{equation}
\gamma^{-}_{\vec n}=\sum_{{\vec m} \in {\rm
B^{(t)}_{NC}}}\Gamma_{{\vec n},\vec{m}}^{\vec{m},{\vec n}}.
\end{equation}

\section{Conclusion}\label{con}

The total rate equation describing a number-conserving population
of the condensate band can be obtained from
Eqs.(\ref{n},\ref{taxaQ},\ref{nscatt},\ref{neva0})

\begin{equation}
\frac{d\textsf{n}_{\vec n}}{dt}=\frac{d\textsf{n}_{\vec
n}}{dt}|_{\rm growth}+\frac{d\textsf{n}_{\vec n}}{dt}|_{\rm
scatt}+\frac{d\textsf{n}_{\vec n}}{dt}|_{\rm evap},
\end{equation}

\noindent where $\textsf{n}_{0}$ and $\textsf{n}_{\vec n}$ denote, respectively, 
the population in the condensate and the population of the elementary
excitations, which are commonly referred to as "thermal cloud".
The equation above gives a complete description for the
thermalization of a system $A$ which is in thermal contact with a
bath $B$. All the information about the dynamics of the
thermalization process is contained in the coefficients $\Gamma$.
Analytical and numerical evaluation of the coefficients for a specific system
can be carried out along the lines described in Refs. \cite{master,pap01}.

The result obtained here can be applied to describe the dynamics of  
sympathetic cooling of a gas in thermal contact with a cooling agent, in terms
of the population in the ground-state and elementary excitations. In 
particular, it can be aplied to the case where the cooling agent thermalize much faster compared
to the thermalization of the system. The description remains valid for the
quantum degenerate regime $T\ll T_{c}$, where $T_{c}$ is the critical 
temperature of the gas. 

The use of the Giradeau-Arnowitt method has opened up the possibility
to describe sympathetic cooling in terms of particles in the
ground state and elementary excitation during the whole cooling process,
i.e., from the case of total absence of particles in the
condensate ($\textsf{n}_0=0$) to the case of a highly populated
ground state ($\textsf{n}_0=N$). In addition, we include the
effects of evaporation which are especially important during the
initial stages of the cooling process. To our knowledge this is
the first time that a formal complete number-conserving
description of sympathetic cooling in terms of particles in the
ground state and elementary excitations is given. 

\section* {Acknowledgments}
ANS thanks fruitful discussions with M. da Mata, 
A.F.R. de Toledo Piza, M.O. da C. Pires, A. Mosk, M. Ameduri, H. Schomerus,
A. Ozorio de Almeida and H.A. Weidenm\"uller. ANS gratefully acknowledges 
M. Weidem\"uller and M.C. Nemes for many discussions and help in
preparing the manuscript.

\end{multicols}
\end{document}